\documentclass[intlimits,twoside,a4paper]{article}

\usepackage[cp1251]{inputenc}

\usepackage[eqsecnum]{cmpj3}

\issue{2018}{21}{4}{43002}
\doinumber{10.5488/CMP.21.43002}
\title[Equilibrium properties of the lattice system with SALR interaction potential]%
{Equilibrium properties of the lattice system with SALR interaction potential on a square lattice:
quasi-chemical approximation versus Monte Carlo simulation}

\author[Ya.G. Groda, V.S. Vikhrenko, D. di Caprio]{Ya.G. Groda\refaddr{label1}, V.S. Vikhrenko\refaddr{label1},
        D. di Caprio\refaddr{label2}}
\addresses{
\addr{label1} Belarusian State Technological University,
13a Sverdlova St., 220006 Minsk, Belarus
\addr{label2} Chimie ParisTech, PSL Research University, CNRS,
Institute of Research of Chimie Paris (IRCP), \\11 Pierre and Marie Curie St.,
75231 Paris, France
}

\date{Received November 2, 2018, in final form November 12, 2018}

\begin{document}

\maketitle

\begin{abstract}
The lattice system with competing interactions that
models biological objects (colloids, ensembles of protein
molecules, etc.) is considered. This system is the lattice fluid on a square lattice with attractive interaction between nearest
neighbours and repulsive interaction between next-next-nearest neighbours. The geometric order parameter is introduced for describing the ordered phases in this system. The critical value of the order parameter is estimated and the phase diagram of the
system is constructed. The simple quasi-chemical approximation
(QChA) is proposed for the system under consideration. The data
of Monte Carlo simulation of equilibrium properties of the model
are compared with the results of QChA. It is shown that QChA
provides reasonable semiquantitative results for the systems studied
 and can be used as the basis for next order
approximations.
\pacs 05.10.Ln, 64.60.De, 64.60.Cn
\end{abstract}

\section{Introduction}

At present, there is a great interest in studying the
processes of self-organization and self-assembly in systems of
a nanoscale range. As the elements of such systems are
supramolecular formations with a sufficiently large molecular
mass, this leads to low velocities of their thermal motion and to 
large, on the molecular scale, characteristic times of the
processes within the system. At the same time, the interaction
between these elements is very complex. Despite their rather
large dimensions, the interactions remain of the same order as
the thermal energy. This leads to a large variety of possibilities
for various phase transitions in such systems at room
temperature.  Examples are solutions of protein
molecules \cite{Nat2004}, clays and soil suspensions \cite{PRL2009},
ecosystems \cite{PRE2015}, etc.

In general, structure elements of such systems attract each
other at small distances due, for example, to the Van der
Waals attraction, and repulse on longer separations because of
the electrostatic interactions (SALR systems, short-range
attractive and long-range repulsive interaction)  \cite{JCP2007,CPL2000}. In the
case of biological molecules, repulsion can also be caused by
elastic deformations of the lipid membranes. In any case, the
attraction between the structural elements of the system
ensures the phase separation, and repulsion --- the formation of
clusters.

One of the simplest methods for investigating the general
properties of SALR systems is to consider their lattice models.
These models are simple enough and allow one to make a detailed analysis
both by analytical methods and computer simulation, using the
Monte Carlo method. A large number of common properties
of such systems can be obtained within these frameworks.

For example, in papers \cite{JCP2014-1,JCP2014-2}, a lattice fluid was studied
with attraction of nearest neighbours and repulsion of the third
neighbours on a plane triangular lattice. Possible
configurations of the ensemble of particles were investigated.
A phase diagram of the system in the mean field
approximation and by Monte Carlo simulation was
constructed. It revealed the existence of several phase
transitions in the system.

In \cite{nap2017} the generalized quasi-chemical approximation (QChA) was
proposed for lattice systems with SALR interaction potential
on a triangular lattice. This approximation has demonstrated
its applicability for estimating the equilibrium properties of
the model in the disordered phase.

In this paper, we present the results for a similar model of
the lattice fluid on a square lattice and propose a geometric
order parameter, which makes it possible to investigate the
ordered phases in the system.

\section{The model and its order parameter}

We consider the lattice fluid consisting of $n$ particles on a
square lattice of $N$ lattice sites. Multiple occupation of sites is
forbidden. Particles that occupy nearest lattice sites and sites
that are neighbours of the third order interact with each other.
The interaction energies are equal to $J_{1}$ and $J_{3}$, respectively. It
is assumed that $J_{1}=-J<0$, and $J_{3}=3J>0$, which corresponds to
the attraction of the nearest neighbours and the repulsion of
the third ones. The second, forth and more distant neighbours
are considered as noninteracting ones.

The simulation of the equilibrium characteristics of the
system under consideration in the grand canonical ensemble
using the Monte Carlo method is performed within the
framework of the Metropolis algorithm \cite{UG1991}. For simulation,
we used a lattice containing $2^{14}$ lattice sites with periodic
boundary conditions. The total length of the simulation
procedure consisted of $70000$ steps of the Monte Carlo
algorithm (MCS). The first $20000$ MCSs were used to
equilibrate the system and were not taken into account at subsequent
averaging.

\begin{figure}[!b]
\centerline{\includegraphics[width=0.85\textwidth]{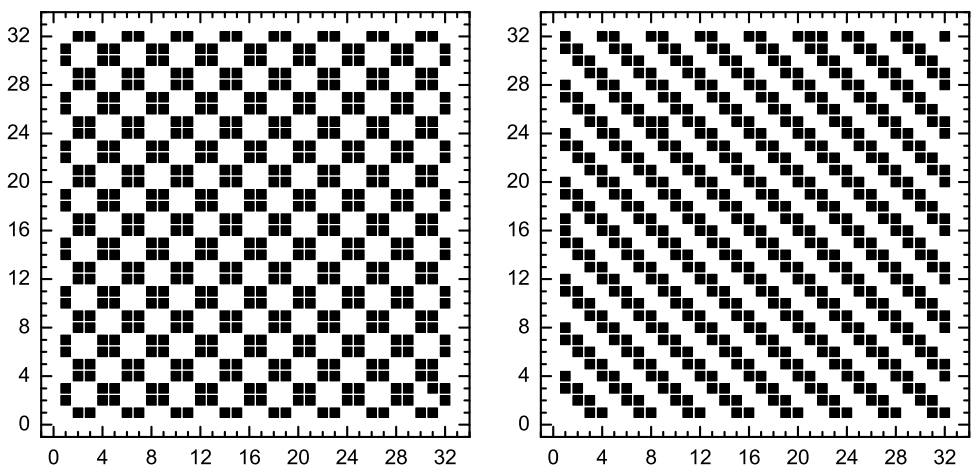}}
\caption{The final screenshot of the system at $\mu=4J$ and $J/k_\text{B}T=1.0$. The left-hand and right-hand parts of the figure correspond to  different launches of the MC-simulation program.} \label{fig1}
\end{figure}

A preliminary simulation on the lattice containing $2^{10}$ lattice sites
has shown that two different types of ordered phases are formed
in the system at low temperatures (below the critical temperature
$T_\text{c}$). The both types of ordered phases are shown in figure~\ref{fig1}.

On smaller lattices, such as those shown in figure~\ref{fig1}, the phases appear
sporadically on different trajectories.
On larger lattices (e.g., of $2^{16}$ lattice sites) the system breaks up
into domains with different types of the ordered structures. The ground state
of the model is degenerated with the energy $-J$ per particle for both configurations. Subsequently, with an increase in temperature,
both states are realized with approximately equal probabilities.

To describe the ordered phases, the initial square lattice
was divided into a system of 8 identical sublattices rotated by
$45^{\circ}$ with the spacing $2a\sqrt{2}$, where $a$ is the lattice spacing of the
initial lattice. In the case of complete ordering of the system at
the lattice concentration $c=0.5$ and at low temperatures, four
sublattices are completely filled ($p$-sublattices) and four
sublattices are completely vacant ($v$-sublattices). This makes it
possible to determine the system order parameter $\delta c$ as the
difference between the particle concentrations on the
sublattices
\begin{align}
\delta{c}=\frac{c_{p}-c_{v}}{2}.
\label{1}
\end{align}
If the sublattices are numbered, we can consider the order parameter matrix
\begin{align}
\delta{c}_{ij}=\frac{|c_{i}-c_{j}|}{2}\,,
\label{1.1}
\end{align}
where $c_{i(j)}$ is the concentration of particles on the sublattice
$i(j)$.

This is a symmetric matrix with the diagonal elements equal to zero and 28 independent non-diagonal elements equal to one or zero for the ordered structures. It is possible to distinguish the ordered phases by the order of filled sublattices or by the structure of the matrix; e.g., even numbers (two or four) of 1 or 0 appear in the sequences in rows and columns of the matrix for squares, while odd numbers (three) are characteristic of stripes. Importantly, for large lattices when both structures are present simultaneously, the scalar order parameter equation~(\ref{1}) distinguishes ordered states from disordered ones.

\begin{figure}[!b]
\centerline{\includegraphics[width=0.45\textwidth]{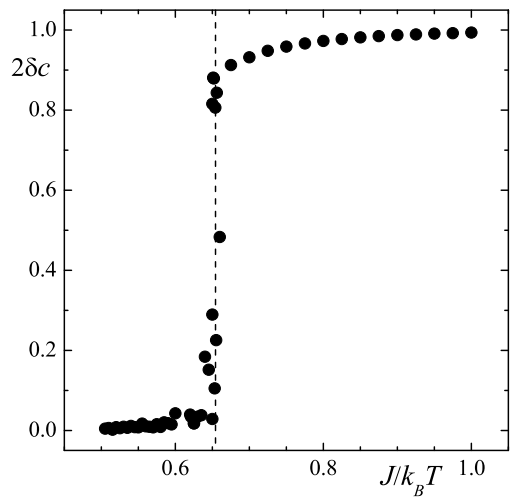}}
\caption{The order parameter versus the inverse temperature at $\mu=4J$ ($c=0.5$).} \label{fig2}
\end{figure}

The order parameter characterizes the strength of the
ordering of the ordered state and is equal to zero in a
disordered state. The total lattice concentration $c$ and the
sublattices concentrations $c_{p}$ and $c_{v}$ are connected by the
expressions (the subscripts 1 and 0 are related to particles and
vacancies, correspondently)
\begin{align}
&c_{p}=c_{1}^{p}=c+\delta{c}, \qquad c_{0}^{p}=1-c_{1}^{p}=1-c-\delta{c},
\label{2}
\\
&c_{v}=c_{1}^{v}=c-\delta{c}, \qquad c_{0}^{v}=1-c_{1}^{v}=1-c+\delta{c},
\label{3}
\\
&c=c_{1}=\frac{n}{N}=\frac{c_{1}^{p}+c_{1}^{v}}{2}\,, \qquad c_{0}=1-c.
\label{4}
\end{align}
The MC simulation shows (see figure~\ref{fig2}) that the order
parameter increases sharply at $J/k_\text{B}T_\text{c}=0.655$ for the chemical
potential $\mu=4J$ which corresponds to the system with the average
concentration $c=0.5$.

Such an increase of the order parameter corresponds to the
order-disorder phase transition, which is of the second order like in the system with the nearest neighbour repulsive interaction \cite{Binder}. In this case, the value $0.655$
can be interpreted as a critical parameter of the system.

Thus, the value of the order parameter can be used to
localize points of the structural phase transitions and to
construct the phase diagram of the model.

\section{The quasi-chemical approximation}

The free energy of the system can be represented as a sum of
the free energy of the reference system $F_\text{r}$ and the
diagrammatic part $F_\text{d}$ \cite{EPJB2003,SSI2005}:
\begin{align}
F=F_\text{r}+F_\text{d}.
\label{5}
\end{align}

The reference system is characterized by the mean
potentials $\phi_{j}^{\,\beta}(n_{i}^{\alpha})$ describing the interaction of a particle
($n_{i}^{\alpha}=1$) or vacancy ($n_{i}^{\alpha}=0$) on the site $i$ of $\alpha$-sublattice with site $j$ of
$\beta$-sublattice. Equation~(\ref{5}) is an identity and the free energy should
not be dependent on the choice of the mean potentials.
Therefore, the mean potentials can be found from the minimal
susceptibility principle \cite{EPJB2000}
\begin{align}
\frac{\partial{F}}{\partial{\phi_{i}^{\alpha}}}=0.
\label{6}
\end{align}

The free energy is a function of the sublattice
concentrations. It is useful to represent it as a function of the
lattice concentration $c$ and the order parameter $\delta{c}$. The latter
can be determined from the extremity condition
\begin{align}
\frac{\partial{F}}{\partial{\delta{c}}}=0,
\label{7}
\end{align}
which is equivalent to the requirement that the chemical
potentials on all the sublattices are equal.

As a first step, one can consider a quasi-chemical
approximation when the diagrammatic part of free energy
contains the two-vertex graph contribution only
\begin{align}
F=\frac{k_\text{B}T}{2}\sum_{\alpha=p}^v\sum_{i=0}^1c_{i}^{\alpha}\left[\ln{c_{i}^{\alpha}}-\sum_{k}z_{k}\ln X_{i}^{{\alpha}(k)}\right]-
\frac{k_\text{B}T}{2}\sum_{k}\frac{z_{k}}{2}\sum_{\alpha,\beta=p}^v\sum_{i,j=0}^1c_{i}^{\alpha(k)}c_{j}^{\,\beta(k)}\left[\frac{W_{ij}^{(k)}}
{X_{i}^{{\alpha}(k)}X_{j}^{{\beta}(k)}}-1\right],
\label{8}
\end{align}
where
\begin{align}
X_{i}^{{\alpha}(k)}=\exp\left[-\beta\phi^k(n_i^\alpha)\right],
\label{9}
\end{align}
$z_k$ is the coordination number for neighbours of $k$-order
($z_1=z_3=4$ for a square lattice).

In this approximation, the diagrammatic part of free energy
is equal to zero and the mean potentials for nearest-neighbours
do not depend on the sublattice structure. The free energy in
the QChA reads
\begin{align}
F^{\text{(QChA)}}(c,\delta c)&=\frac{k_\text{B}T}{2}\sum_{i}c_i\left[\ln (c_i^2-\delta c^2)-2z_1\ln X_i\right]-
\frac{k_\text{B}T}{2}z_3\left(\ln{Z_0^pZ_0^v}+c\ln{\xi_v\xi_p}\right)
\nonumber\\
&+\frac{k_\text{B}T}{2}\delta c\left(\sum_k\ln{\frac{c_i+\delta c}{c_i-\delta c}}-z_3\ln\frac{\xi_p}{\xi_v}\right),
\label{10}
\end{align}
where
\begin{align}
&W=\exp\left(-\frac{J_1}{k_\text{B}T}\right); \qquad \Omega=\exp\left(-\frac{J_3}{k_\text{B}T}\right);
\label{11}
\end{align}
\begin{align}
&\eta=-\frac{c_1-c_0}{2c_0}+\sqrt{\left(\frac{c_1-c_0}{2c_0}\right)^2+\frac{c_1}{c_0}W};
\label{12p}
\\
&X_0=\sqrt{c_0+\frac{c_1}{\eta}}\,, \qquad X_1=\eta X_0;
\label{12}
\\
&\xi_{p(v)}=-\frac{c_1-c_0\pm 2\Omega\delta c}{2(c_0\mp\delta c)}+\sqrt{\left[\frac{c_1-c_0\pm 2\Omega\delta c}{2(c_0\mp\delta c)}\right]^2+\frac{c_1\pm\delta c}{c_0\mp\delta c}\Omega};
\label{13}
\\
&Z_0^vZ_0^p=c_0^v+\frac{c_1^v}{\xi_v}=c_0^p+\frac{c_1^p}{\xi_p}.
\label{14}
\end{align}

All the thermodynamic characteristics can be investigated
with equation~(\ref{10}) for the free energy. Thus, the chemical
potential $\mu$, the thermodynamic factor $\chi_T$ and the correlation
function $g_k(1;1)$ for two nearest- and next-next-nearest
neighbours (at $k=1$ and $3$, respectively) are determined by the
expressions
\begin{align}
&\beta\mu=\left(\frac{\partial (\beta F)}{\partial c}\right)_T,
\label{15}
\\
&\chi_T=\frac{\partial (\beta\mu)}{\partial\ln c}\,,
\label{16}
\\
&g_k(1;1)=\frac{2}{z_kc^2}\left(\frac{\partial F}{\partial J_k}\right)_T.
\label{17}
\end{align}

\section{Calculation and simulation results}

\begin{figure}[!b]
\centerline{\includegraphics[width=0.45\textwidth]{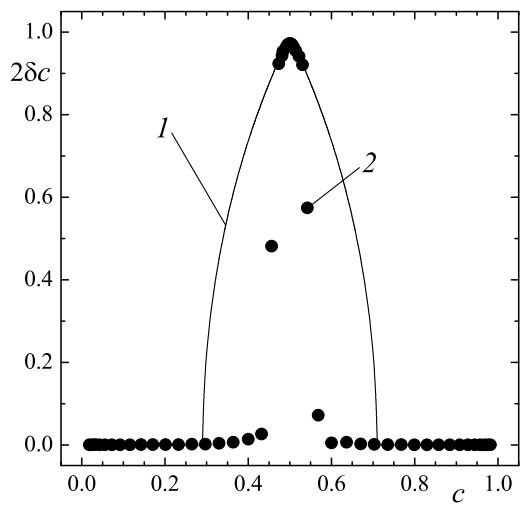}}
\caption{The order parameter versus concentration at $\beta J =0.8$. The solid line \textit{1} represents the QChA results, the
full circles \textit{2} are the MC simulation data.} \label{fig3}
\end{figure}

The most important structural feature of an ordered state is
the order parameter. The comparisons of the calculation and
simulation results for the order parameter are shown in figure~\ref{fig3}.

\begin{figure}[!t]
\centerline{\includegraphics[width=0.45\textwidth]{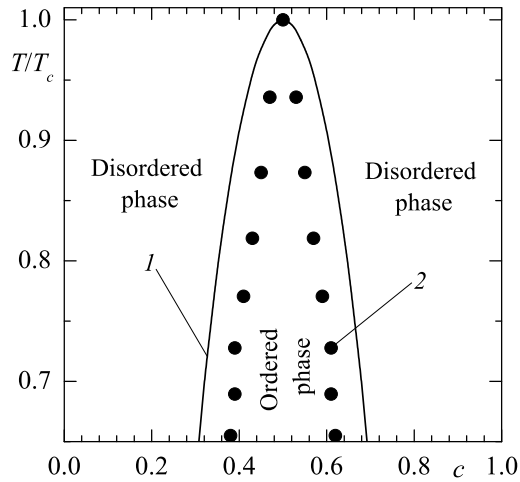}}
\caption{The order-disorder phase transition curves. The solid line \textit{1} represents the QChA results, the
full circles \textit{2} are the MC simulation data.} \label{fig4}
\end{figure}

The order parameter is used to determine the phase
transition curve. The corresponding phase diagram is
represented in figure~\ref{fig4}. One can note that QChA leads to a wider
area of the ordered phase in the system as compared to the
Monte Carlo simulation results. In addition, the critical
temperature is overestimated by approximately 30\% in this
approximation. These deviations are known to be typical of
QChA \cite{Domb,Huang}.

It should be noted that the phase diagram of the system with the competing interaction on a square lattice is much simpler than that of the system on a triangular lattice \cite{JCP2014-2} which is probably the consequence of a close-packed structure of the latter.

\begin{figure}[!t]
\centerline{\includegraphics[width=0.45\textwidth]{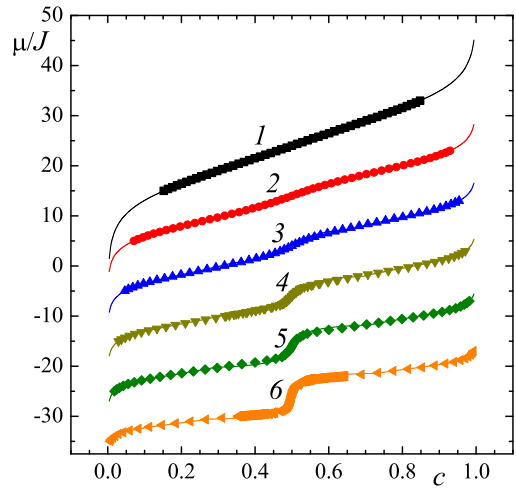}}
\caption{(Colour online) The chemical potential (in units of the nearest neighbour interaction energy $J$) versus concentration at $\beta J =0.3$ (\textit{1}); 0.5 (\textit{2}); 0.6 (\textit{3}); 0.7 (\textit{4}); 0.8 (\textit{5}) and 1.0 (\textit{6}). The solid lines represent the QChA results, the full circles are the MC simulation data. Each group of curves is shifted down by 10 units along the $\mu$ axis with respect to the previous one for better visibility. The unshifted curve (\textit{3}) is characterized by $\mu /J=4$ at $c=0.5$, and this point is the same for all the temperatures. Thus, the groups of curves (\textit{1}) and (\textit{2}) are shifted up from their true position, while (\textit{4}), (\textit{5}) and (\textit{6}) are shifted down. The same curves numbers are kept in figures~\ref{fig6}--\ref{fig8}.} \label{fig5}
\end{figure}

\begin{figure}[!t]
\centerline{\includegraphics[width=0.45\textwidth]{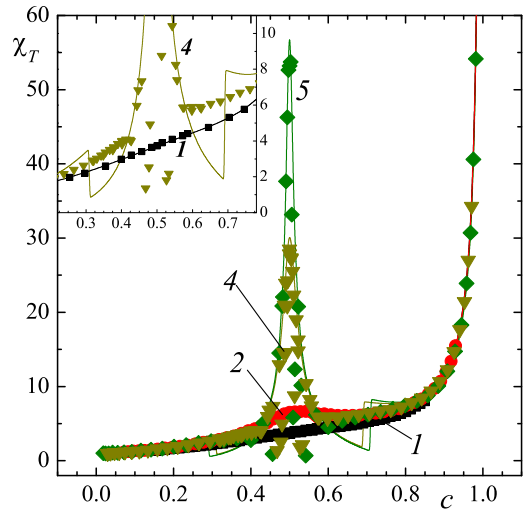}}
\caption{(Colour online) The thermodynamic factor versus concentration at $\beta J =0.3$ (\textit{1}, square); 0.5 (\textit{2}, circle); 0.7 (\textit{4}, down triangle) and 0.8 (\textit{5}, diamond). The solid lines represent the QChA results, the symbols are the MC simulation data.} \label{fig6}
\end{figure}

The chemical potential isotherms are shown in figure~\ref{fig5}. The
ordered phase exists at temperatures below critical ($\beta J=0.6$,
$0.7$, $0.8$ and $1.0$) where a steep increase of the chemical
potential with an increase of the particle concentration is
observed. The concentration derivative of the chemical
potential or the thermodynamic factor (\ref{16}) indicates the
second-order phase transition by discontinuities at the
concentrations that correspond to the phase transition points.
The strong peaks at $c=0.5$ correspond to the most ordered
states of the system at corresponding temperatures (see figure~\ref{fig6}).

The thermodynamic factor is inverse to the concentration
fluctuations. The latter grow immediately after the transition
from a disordered to an ordered state and then decrease
systematically until concentration $0.5$ is reached when the
most ordered state is possible. Physically, this means that
concentration fluctuations are suppressed in the ordered states.

In QChA, the chemical potential and the thermodynamic
factor were calculated by numerical differentiating of the free
energy expression~(\ref{10}). However, such a differentiation of the
chemical potential extracted from MC simulations may not be
used. In this case, the thermodynamic factor can be calculated
as the value inversely proportional to the mean square
concentration fluctuations
\begin{align}
\chi_T=\frac{\langle n \rangle}{\langle (n-\langle n \rangle)^2 \rangle}.
\label{18}
\end{align}
The parameter $\chi_T$ plays an important role in describing the diffusion process in lattice fluids \cite{Phys2001}.

\begin{figure}[!b]
\centerline{\includegraphics[width=0.45\textwidth]{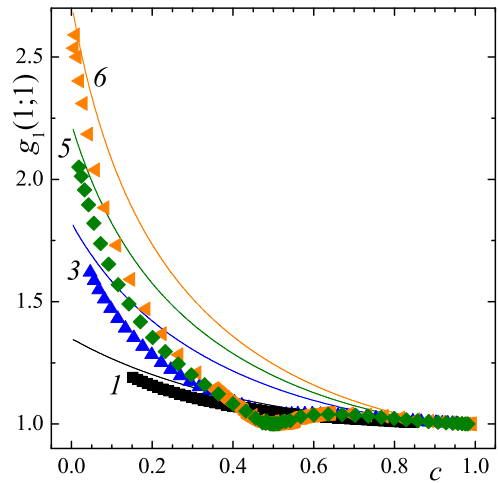}}
\caption{(Colour online) The correlation functions for nearest neighbours versus concentration at $\beta J =0.3$ (\textit{1}, square); 0.6 (\textit{3}, up triangle); 0.8 (\textit{5}, diamond) and 1.0 (\textit{6}, left-hand triangle). The solid lines represent the QChA results, the symbols are the MC simulation data.} \label{fig7}
\end{figure}

The calculation and simulation results for the chemical
potential and thermodynamic factor are in a good agreement
except for those at close vicinity of the second order phase
transition curve where the QChA curves show abrupt jumps.

A short range ordering is characterized by the correlation
functions~(\ref{17}). These functions are more informative objects
than the distribution functions because they represent the
deviation of the short range correlations in interacting systems
from the case of noninteracting (Langmuir) lattice gases where
they are equal to one.

\begin{figure}[!t]
\centerline{\includegraphics[width=0.45\textwidth]{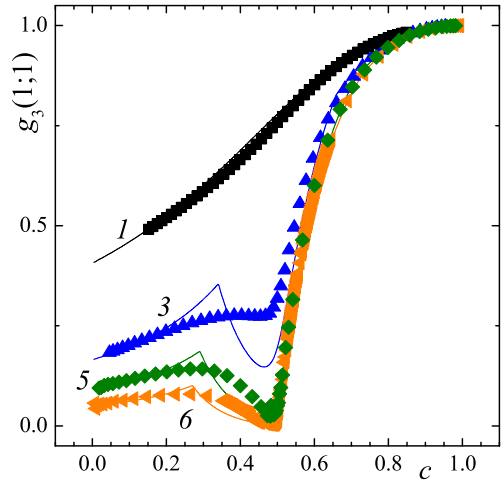}}
\caption{(Colour online) The correlation functions for the next-next-nearest neighbours versus concentration at $\beta J =0.3$ (\textit{1}, square); 0.6 (\textit{3}, up triangle); 0.8 (\textit{5}, diamond) and 1.0 (\textit{6}, left-hand triangle). The solid lines represent the QChA results, the symbols are the MC simulation data.} \label{fig8}
\end{figure}

The correlation functions for nearest and next-next-nearest
neighbours are show in figures~\ref{fig7} and \ref{fig8}, respectively.

The long range ordering manifests itself in a short range
structure as well. At temperatures $T<T_\text{c}$ (see curves \textit{3}, \textit{5} and
\textit{6}) the probability to find two next-next-nearest neighbour sites
occupied by particles becomes very low. Of course, the higher is
the temperature, the less pronounced difference from the
Langmuir gas behaviour is observed.

The simulation and analytical results satisfactorily match
each other in the disordered phase. They significantly differ at
intermediate concentrations and at low temperatures due to the
problems concerning the determination of the order
parameter and the critical temperature in QChA. The most
significant differences appear for the correlations on the
nearest neighbour lattice sites.

\section{Conclusion}

The lattice system with an attractive interaction between
nearest neighbours and repulsive interaction between
next-next-nearest neighbours has been studied.

It is shown that the competing interactions lead to the order-disorder phase transitions. The order parameter $\delta c$ is used as the
indicator of the second order phase transitions. With this
parameter, it was established that the critical value of the
interaction parameter is equal to $|J_1|/k_\text{B}T_\text{c}=0.655\pm0.005$, and the
phase diagram of the system was constructed.

The quasi-chemical approximation is found to be self-consistently constructed through the mean potentials that
describe the interaction of a particle or a vacancy with its
nearest and next-next-nearest neighbour lattice sites. The
chemical potential, thermodynamic factor and correlation
functions are determined both in the QChA and in the Monte
Carlo simulations. The chemical potential demonstrates
irregular behaviour in the phase transition region, while the
thermodynamic factor indicates strong suppressing of
fluctuations  that are inherent to the ordered states of the system.
The complicated behaviour of the correlation functions that
reflects structural peculiarities of the system demonstrate a
great importance of competing interactions.

The order parameter of the system $\delta c$ is determined in the
QChA with significant errors. This leads to errors in
determining the critical temperature of the system, which is
overestimated by approximately  30\%. As a result, the
quasichemical approximation fails to reproduce the structural
characteristics of the system with competing interactions. At
the same time, the thermodynamic characteristics such as the
chemical potential and the thermodynamic factor are
determined in the quasi-chemical approximation with a
sufficiently high accuracy.

Thus, the developed approach allows us to correctly
describe the qualitative features of the structural properties of
the systems with competing interactions, and can be used to
quantify the thermodynamic characteristics of these systems.

\section*{Acknowledgements}
The project has received funding from the European Union's Horizon 2020 research and
innovation programme under the Marie Sk\l odowska-Curie grant agreement No 73427,
Institute for Nuclear Problems of Belarusian State University (agreement No 209/103) and the Ministry of Education of Belarus.

\newpage
\ukrainianpart

\title{Рівноважні властивості ґраткової системи з потенціалом взаємодії типу  SALR на квадратній ґратці: квазiхiмiчне наближення у порівнянні з симуляцiями Монте Карло}
\author{Я.Г. Грода\refaddr{label1}, В.С. Віхренко\refaddr{label1}, Д. ді Капріо\refaddr{label2}}
\addresses{
\addr{label1} Білоруський державний технологічний університет,
вул. Свєрдлова, 13a, 220006 Мінськ, Білорусь
\addr{label2} Дослiдницький унiверситет науки та лiтератури Парижу, ChimieParisTech — CNRS, Iнститут хiмiчних дослiджень Парижу, Париж, Францiя
}

\makeukrtitle

\begin{abstract}
Розглядається ґраткова система з конкурентними взаємодіями, що моделює біологічні об'єкти (колоїди, ансамблі протеїнових молекул і т.д.). 	
Ця система представляється ґратковим плином на квадратній ґратці з притягальною взаємодією між найближчими сусідами і відштовхувальною взаємодією між наступними  за наступними до найближчих сусідів. Для опису впорядкованої фази в такій системі вводиться геометричний параметр порядку. Зроблена оцінка критичного значення параметра порядку і побудована фазова діаграма системи. Для системи, що розглядається,  запропоновано просте квазіхімічне наближення
(КХН). Дані симуляцій  Монте Карло рівноважних властивостей моделі порівнюються з результатами  КХН. Показано, що  КХН забезпечує прийнятні напівкількісні результати для системи, що вивчається, і може використовуватися як базис для наступних наближень.
\keywords ґраткова модель плину, конкуруюча взаємодія, фазовий перехід порядок-безлад, симуляції Монте Карло, квазіхімічне наближення, фазова діаграма
\end{abstract}

\end{document}